\documentclass[aps,prl,twocolumn,superscriptaddress,longbibliography]{revtex4-1}

\usepackage[colorlinks=true,citecolor=blue]{hyperref} 
\usepackage{color}
\usepackage{float}
\usepackage{dsfont}
\usepackage{amsmath}
\usepackage{amssymb}
\usepackage{graphicx}
\usepackage{babel}

\usepackage{xcolor}

\usepackage{graphicx}
\usepackage{dcolumn}
\usepackage{bm}

\begin{document}

\title{Electric field control of moir\'e skyrmion phases in twisted multiferroic NiI$_2$ bilayers}

\author{Tiago V. C. Ant\~ao}
\affiliation{Department of Applied Physics, Aalto University, 02150 Espoo, Finland}

\author{Jose L. Lado}
\affiliation{Department of Applied Physics, Aalto University, 02150 Espoo, Finland}

\author{Adolfo O. Fumega}
\affiliation{Department of Applied Physics, Aalto University, 02150 Espoo, Finland}

\date{\today}

\begin{abstract}
Twisted magnetic van der Waals materials provide a flexible platform to engineer new forms of unconventional magnetism. Here we demonstrate the emergence of electrically tunable topological
moir\'e magnetism in twisted bilayers of the spin-spiral multiferroic NiI$_2$. We establish a rich phase diagram featuring uniform spiral phases, a variety of $k\pi$-skyrmion lattices, and nematic spin textures ordered at the moir\'e scale. The emergence of these phases is driven by the local stacking and the resulting modulated frustration in the spin spiral stemming from the moir\'e pattern. Notably, when the spin-spiral wavelength is commensurate with the moir\'e length scale by an integer $k$, multi-walled skyrmions become pinned to the moir\'e pattern. 
We show that the strong magnetoelectric coupling displayed by the moir\'e multiferroic allows the electric control of the $k\pi$-skyrmion lattices by an out-of-plane electric field, which couples to the moir\'e-induced electric polarization. While adiabatic changes in the electric field preserve the topology of the spin configurations, abrupt variations can trigger transitions between different skyrmion lattice ground states. Our results establish a highly tunable platform for skyrmionics based on twisted van der Waals multiferroics, potentially enabling a new generation of ultrathin topologically-protected spintronic devices.

\end{abstract}

\maketitle

\paragraph*{Introduction.-}

Skyrmions are topologically protected magnetic structures that have garnered significant attention due to their potential applications in spintronics, particularly in the areas of data storage and manipulation  \citep{Fert_2013, Back_2020, Wang_2022_skyrmions, Zhang_2023}. These nanoscale spin textures are of great interest for advanced energy-efficient memory technologies because of their stability and unique spin configurations. Skyrmions have been observed or proposed in a variety of systems, including two-dimensional (2D) magnets  \citep{Kezsmarki_2015, Powalla_2023, Zhang_2023_generation, Yao_2023}, magnetic thin films  \citep{Desautels_2019}, and material interfaces  \citep{Li_2014}. Their formation is typically driven by mechanisms such as the Dzyaloshinskii-Moriya (DM) interaction  \citep{Dzyaloshinsky_1958, Moriya_1960}, magnetic frustration  \citep{Muhlbauer_2009, Yuan_2023, Yu_2010, Okubo_2012, Shen_2023}, and other types of magnetic anisotropy modulation  \citep{Stavrou_2021}. Despite the excitement surrounding skyrmions, their emergence in insulating materials is relatively rare \citep{Seki_2012,Amoroso_2020,Das_2024}. This presents a challenge for integrating skyrmions into insulating platforms, which are particularly attractive for low-power spintronic applications.

\begin{figure}[t!]
    \centering
    \includegraphics[width=\columnwidth]{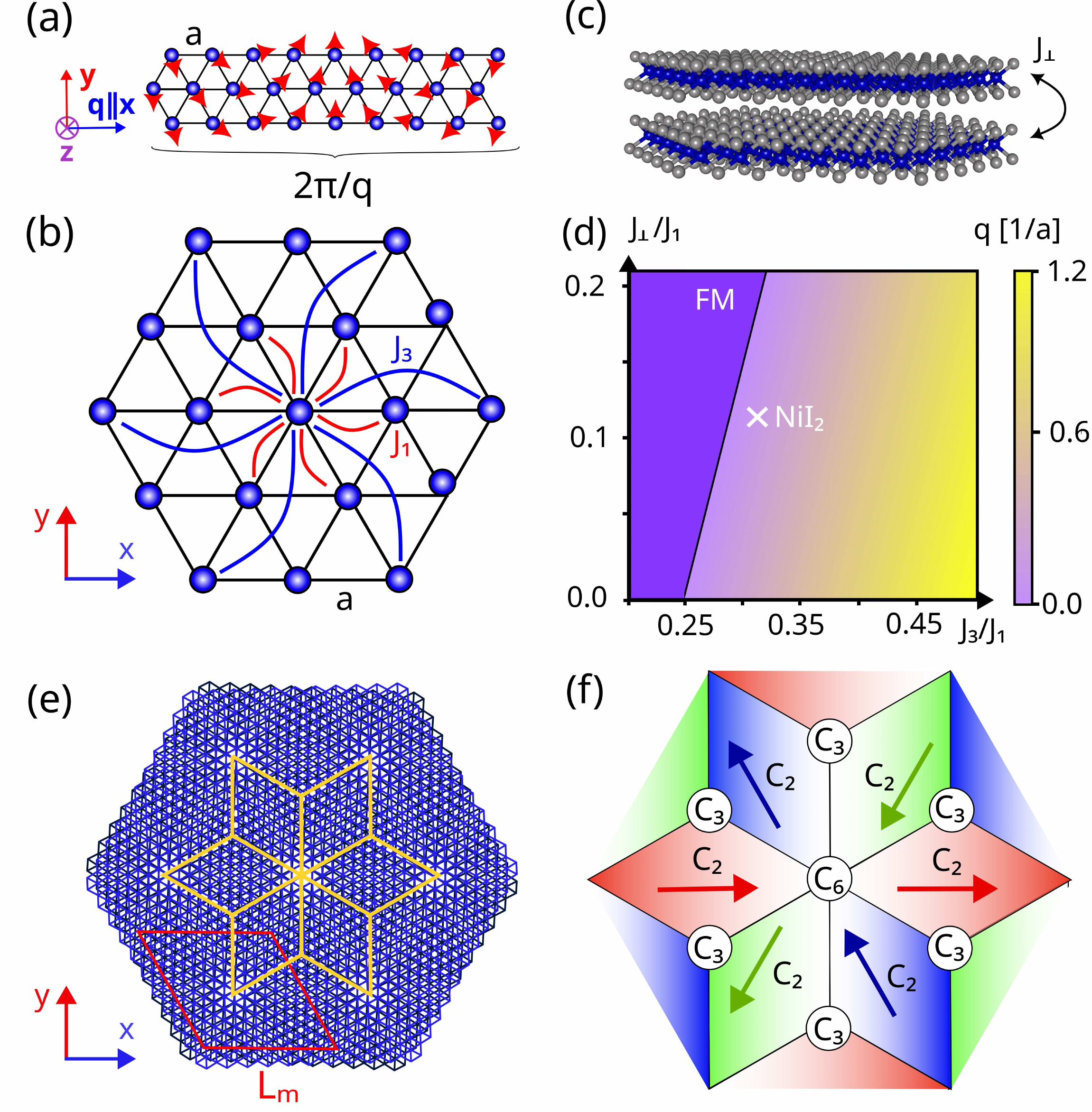}
    \caption{(a) Schematic of the spin-spiral of monolayer NiI$_2$ with wavelength $2\pi/q$.(b) Exchange interactions $J_1$ and $J_3$ of the Ni triangular lattice leading to the non-collinear spin-spiral state.  (c) Schematic of a NiI$_2$ twisted bilayer, with Ni atoms represented in blue, and I atoms in grey. (d) Magnitude of the propagation vector $q$ as a function of the interlayer exchanges $J_\perp/J_1$ and $J_3/J_1$.  (e) Moir\'e pattern of twisted bilayer NiI$_2$, showing the moir\'e unit cell (red),
    and the spatial patches with $C_2$ symmetric rhombohedral stacking along different directions (yellow). (f) Schematic of the domains of $q$ originated by the local broken symmetry.}
    \label{fig:schematic}
\end{figure}

Two-dimensional van der Waals (vdW) materials provide an exciting new avenue for engineering exotic quantum phases of matter \citep{Geim2013}, including the design of
exotic magnetic phases\citep{Burch2018,Gibertini2019,Blei2021}. 
The weak vdW bonding between the layers in this class of materials allows easily reach their monolayer limit, establishing a rich family of 2D building blocks  \citep{Cui_2018, Yuan_2019,Ugeda2016,Fei_2018,Huang_2018,Lee_2016,Gong_2017,Zhang_2019,Posey_2024}. 
The monolayers can be stacked together to generate heterostructures with emergent collective states \citep{Song_2021,Vano_2021,Kezilebieke_2020}.
Moreover, a twist angle can be introduced between stacked layers, giving rise to a moir\'e length scale that drives the emergence of complex phase diagrams  \citep{Andrei2021,Nuckolls_Yazdani_2024}. For example, twisting has been shown to induce multiferroicity and skyrmionic patterns in bilayer chromium trihalides (CrX$_3$, X = Br, I, Cl) \citep{Fumega_2023,Akram2021,PhysRevB.108.L100401,PhysRevB.108.174440,Xie2023,Kim_2024} and transition metal dichalcogenides  \citep{Haavisto_2022,PhysRevLett.133.026801}. Additionally, exotic magnetic orders, including whirls and stripes, have been 
proposed in twisted 2D magnets like RuCl$_3$  \citep{Akram_2024}. The isolation of monolayer NiI$_2$\citep{Song_2022,Amini_2024}, has recently provided a new avenue for engineering matter by introducing multiferroic behavior as a new building block within magnetic vdW materials \citep{Burch_2018, Wang_2022, Liu_2023}. Multiferroic materials exhibit simultaneous electric and magnetic order  \citep{Spaldin_2007, Fiebig_2016}. In particular, NiI$_2$ is a type-II multiferroic whose ferroelectricity originates from its helical magnetic order and strong spin-orbit coupling  \citep{Fumega_2022} resulting in a strong magnetoelectric coupling \citep{Kurumaji_2013,Amini_2024}. 
It has been reported that the multiferroic order in NiI$_2$ can be tuned by external factors 
such as strain  \citep{Han_2020}, pressure  \citep{Kapeghian_2024, Liu_2024_pressure}, substrate engineering  \citep{PhysRevB.109.195422} or cobalt substitution  \citep{doi:10.1021/acs.chemmater.4c01053}. 
However, the emergence of exotic magnetic orders displaying strong magnetoelectric coupling in NiI$_2$ moir\'e heterostructures has remained unexplored.

In this letter, we 
demonstrate that twisted bilayer NiI$_2$ allows engineering topological magnetic
textures, realizing a whole new family of multiferroic orders. 
We show that the moir\'e created by the twist angle enhances the frustration in the system and produces a variety of exotic magnetic orders, specifically exotic $k\pi$-skyrmions driven by the commensurability of the spin spiral wavelength and moir\'e length scale. Additionally, we show that the multiferroic order of twisted NiI$_2$ provides an unprecedented level of control over skyrmion lattice states using external electric fields, allowing to drive the system between different skyrmionic configurations. We provide a phase diagram for the skyrmion phases in the twisted bilayer and demonstrate the electric-field tunability of these magnetic states, establishing twisted bilayer NiI$_2$ as a versatile platform for engineering and manipulating topological spin textures.

\begin{figure}
    \centering
    \includegraphics[width=\columnwidth]{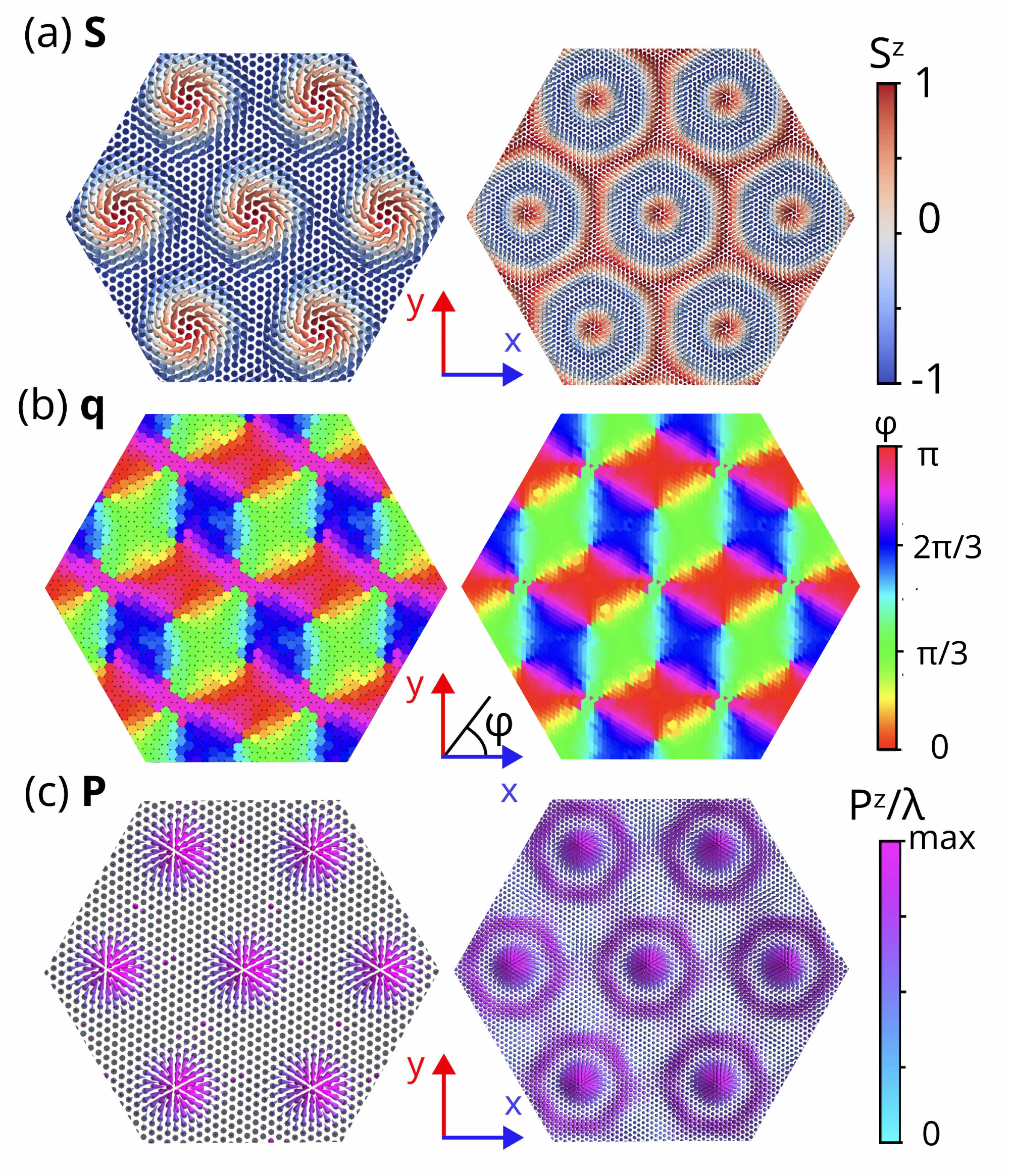}
    \caption{(a) Spin $\boldsymbol{S}$, (b) propagation vector $\boldsymbol{q}$, and (c) polarization $\boldsymbol{P}$ of a skyrmion and skyrmionium lattice in twisted bilayer NiI$_2$. These occur for twist angles $\theta\approx3.89^\circ$ and $\theta\approx2.13^\circ$, left and right panels respectively.
    The panels show the top layer of a twisted bilayer, with the bottom one featuring an analogous
    profile.}
    \label{fig:Ss_qs}
\end{figure}

\paragraph*{Twisted bilayer NiI$_2$ model-}
In monolayer NiI$_2$, multiferroicity emerges from the coexistence of a non-collinear magnetic spin-spiral order and strong spin-orbit coupling  \citep{Fumega_2022,Riedl_2022} via the inverse Dzyaloshinskii-Moriya (DM) effect  \citep{Katsura_2005, Mostovoy_2006, Jia_2006}. This effect causes a non-centrosymmetric displacement of electric charge due to the non-collinear spin structure leading to an electric polarization. The magnetic ground state of monolayer NiI$_2$ corresponds to a co-planar spin spiral magnetic order
(Fig.~\ref{fig:schematic}(a)). The spin spiral phase arises from the magnetic frustration induced by the competition between ferromagnetic first-neighbor $J_1$ and antiferromagnetic third-neighbors $J_3$ magnetic exchange interactions  \citep{Sodequist_2023,Amini_2024,2024arXiv240721645R} (Fig.~\ref{fig:schematic}(b)). In the bulk limit, additional interactions, such as Kitaev anisotropic interaction and biquadratic exchange, have also been proposed to influence the ground state spin configuration in NiI$_2$  \citep{Li_2023}. However, in the ultrathin limit, theoretical analyses have found these terms to be negligible \citep{PhysRevB.109.195422}, and atomic-scale experiments have shown that a minimal classical Heisenberg $J_1-J_3$ model suffices to accurately describe the observed ground state  \citep{Amini_2024}. 
The magnetization in the spin-spiral phase of monolayer NiI$_2$ is described by the propagation vector $\boldsymbol{q}$, determining the direction and wavelength $2\pi/q$ of the spin-spiral.
The $\boldsymbol{q}$ vector is intrinsically constrained by the exchange interactions of the spin Hamiltonian to the direction of the first (or third) neighbors and its periodicity is determined by the $J_3/J_1$ ratio \citep{PhysRevB.93.184413,Glittum_2021}. In particular,  $J_3\approx-0.3J_1$ is found to reproduce the experimentally observed spiral periodicity  \citep{Amini_2024}. 
In addition, experimentally, the magnetization is observed to be mostly in-plane  \citep{Amini_2024}, which 
is accounted for by an in-plane anisotropy term $A_z=-0.02J_1$, which fixes the rotation plane as $\hat{\boldsymbol{e}}=\hat{\boldsymbol{z}}$, as depicted in Fig.~\ref{fig:schematic}(a).
When two NiI$_2$ layers are stacked an antiferromagnetic interlayer exchange $J_{\perp,(ij)}$ is established (Fig.~\ref{fig:schematic}(c)). $J_{\perp,(ij)}$ is parameterized by an exponentially decaying function peaking at the next nearest interlayer neighbor distance at zero twist angle with strength $J_\perp$. 
The total Hamiltonian for twisted NiI$_2$ bilayer reads

\begin{align}\label{eq:modeltwistednii2}
H=&J_1 \sum_{\left<i,j\right>_1}\boldsymbol{S}_i\cdot\boldsymbol{S}_j+J_3 \sum_{\left<i,j\right>_3}\boldsymbol{S}_i\cdot\boldsymbol{S}_j\nonumber\\&+ A_z\sum_i \left(S^z_i\right)^2+\sum_{\left<i,j\right>_\perp} J_{\perp,(ij)} \boldsymbol{S}_i\cdot\boldsymbol{S}_j,
\end{align}

where $\left<\cdot,\cdot\right>_n$ denotes the $n$th nearest neighbors and $\left<\cdot,\cdot\right>_\perp$ refers to summation over sites in distinct layers. 
The emergence of distinct orders in the twisted bilayer arises from the additional frustration introduced by the interlayer coupling (Fig.~\ref{fig:schematic}(c)). 
Note that, as opposed to other twisted magnets  \citep{Fumega_2023,Akram2021,Akram_2024}, the sign of $J_\perp$ does not have a stacking dependence, and the emerging orders in this system will be driven by the interplay between the spin-spiral and moir\'e length scales. 
Physically, the presence of antiferromagnetic interlayer interactions causes a local anti-alignment of spins on the top and bottom layers resulting in the two layers having the same direction and magnitude of $\boldsymbol{q}$. The first effect of  Fig. \ref{fig:schematic}(d)  transition from a ferromagnetic phase to a spin-spiral phase at a ratio of $J_3/J_1=0.25$ when $J_\perp=0$, and this critical value is enhanced for $J_\perp>0$. 
The value of the ratio $J_3/J_1$ for twisted bilayer NiI$_2$ is taken to be approximately the same as the monolayer, and the interlayer exchange is estimated to be given by 
$J_\perp/J_1\approx-0.14$ \citep{Kurumaji_2013, Li_2023}, marked in the phase diagram Fig.~\ref{fig:schematic}(d).

When the two layers are twisted, the phase diagram of Fig.~\ref{fig:schematic}(d) is found to be unchanged up to twist angles of $6^\circ$, so the renormalization of the magnitude of $\boldsymbol{q}$ does not depend on the angle for small twist angles. However, the moir\'e pattern induces regions of higher spin density along certain spatial directions within each moi\'e unit cell, as illustrated in Fig.~\ref{fig:schematic}(e). In Fig.~\ref{fig:schematic}(f), the different colored regions correspond to areas of the moir\'e pattern where the rotational symmetry of the crystalline environment is broken from the monolayer $C_6$ down to a $C_2$ rhombohedral stacking, promoting the emergence of patches with well defined $q$-vector directions. In the center of the moir\'e supercell, a $C_6$ symmetric monoclinic stacking area remains, where six $q$-vector patches coalesce. Additional $C_3$ symmetric regions appear, surrounded by three $q$-vector patches. This local modulation of $\boldsymbol{q}$ into a function of position $\boldsymbol{q}_i$ is, contingent on the moir\'e supercell being large enough for these symmetries to approximately hold, as well as $J_\perp$ being strong enough relative to the moir\'e unit cell size $L_m$ to generate a sufficiently strong pinning potential. 
Therefore, a finite twist angle between layers will introduce an extra moir\'e length scale $L_m$ that will interplay with the spin spiral length scale $\boldsymbol{q}$ of each layer, thus giving rise to exotic magnetic scales.

\begin{figure}
\begin{centering}
\includegraphics[width=\columnwidth]{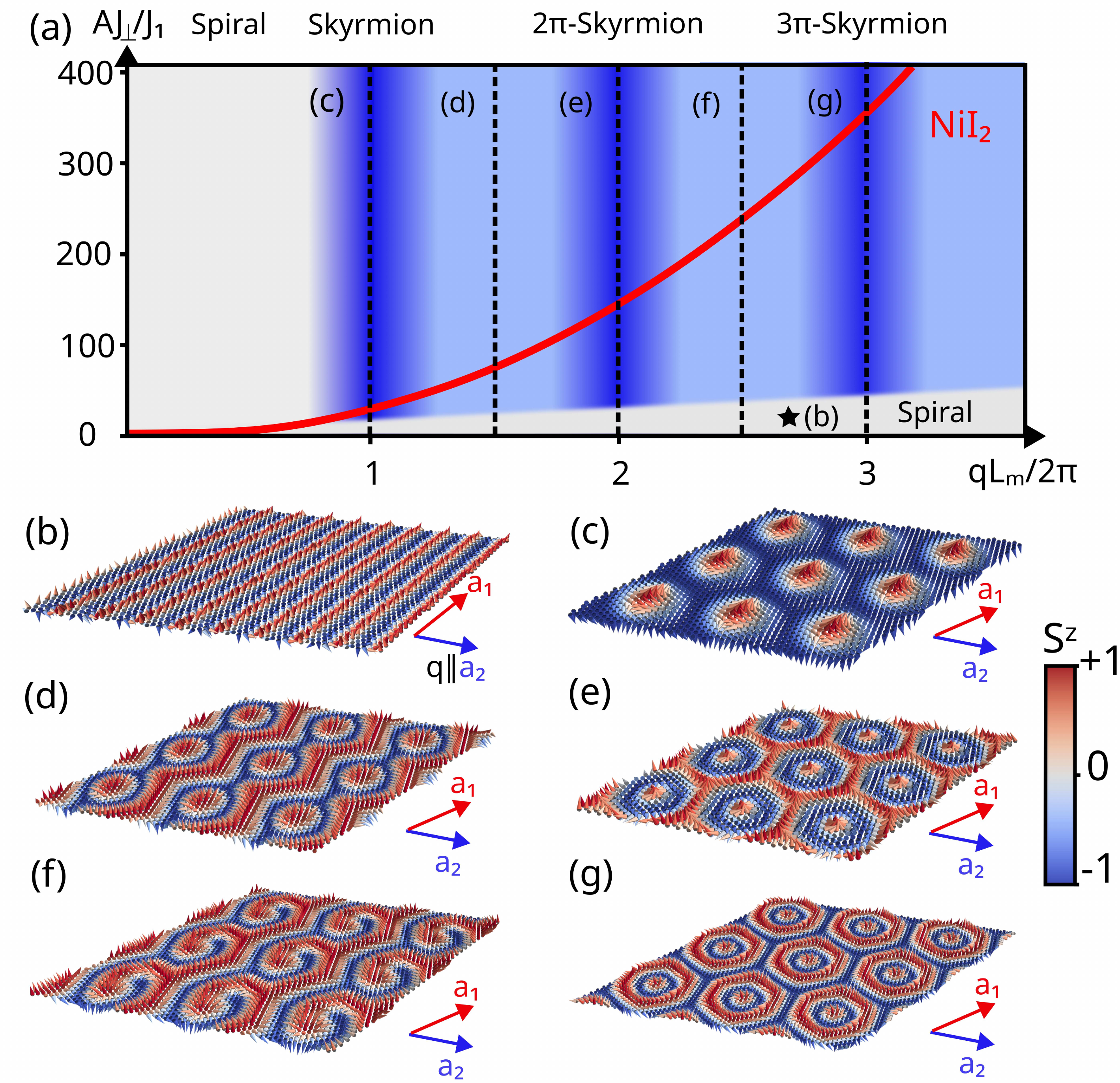}\par\end{centering}
\caption{(a) Phase diagram of twisted bilayer NiI$_2$ as a function of normalized interlayer exchange $AJ_\perp/J_1$ and $qL_m/2\pi$. 
The factor $A=L_m^2/a^2$ introduces the relative area between the moir\'e and the single unit cells, ensuring that the same ground states are represented along vertical lines in the phase diagram.
The red line represents NiI$_2$ as $L_m$ increases or, equivalently, as the twist angle becomes smaller. (b-g) Representative ground state solutions in the phase diagram of the panel (a): (b) uniform spin spiral, (c), (e), and (g) display the simple, $2\pi-$ and $3\pi-$skyrmion lattices respectively., and (d) and (f) show intermediate phases with nematic order between commensurate solutions. 
Only the top layer of the twisted bilayer is shown for clarity.
}
\label{fig:phase_diagram}
\end{figure}

\paragraph*{Phases in twisted NiI$_2$ bilayer-}

We move on to computing the magnetic ground states arising from the model for twisted NiI$_2$ as a function of the competing spin spiral and moir\'e length scales. 
We find the lowest energy
configurations under generalized twisted boundary conditions,
which allows us to account for ground states where the spiral wavelength and moir\'e length scale are not commensurate. 
Two paradigmatic examples of the kind of topological magnetic phases achievable in twisted bilayer NiI$_2$ are the skyrmion and skyrmionium  (2$\pi$-skyrmion) lattices \citep{Powalla_2023,Kolesnikov_2018,Wang_2020_skyrmionium,Yang_2023_conversion}. 
These are plotted in Fig.~\ref{fig:Ss_qs}(a) 
considering the commensurate cases $qL_m/2\pi=1$ and $qL_m/2\pi=2$, with twist angles $\theta\approx3.89^\circ$ and $\theta\approx2.13^\circ$ respectively. 
Associated with these magnetic textures, in Fig.~\ref{fig:Ss_qs}(b) we have computed the local $\boldsymbol{q}$. The results match well with the schematic presented in Fig.~\ref{fig:schematic}(f), confirming that the moir\'{e} length scale induces the pinning of the $\boldsymbol{q}$ vector directions. 
Finally, the presence of strong spin-orbit coupling $\lambda$ induces an emergent electric polarization $\boldsymbol{P}_{ij}=\lambda \boldsymbol{r}_{ij}\times\left(\boldsymbol{S}_{i}\times\boldsymbol{S}_{j} \right)$ in non-collinear magnets \citep{Katsura_2005,Mostovoy_2006}. 
Figure~\ref{fig:Ss_qs}(c) shows the emergent polarization associated with the skyrmion and skyrmionium lattices in twisted NiI$_2$. 
It can be seen that the emergent polarization points mostly in the out-of-plane direction
$\boldsymbol{z}$. This starkly contrasts the monolayer case and is a key feature that allows the twisted bilayer to couple to out-of-plane electric fields.

The ground-state solutions as a function of the twist angle are presented
in the phase diagram shown in Fig.~\ref{fig:phase_diagram}(a).
Variations of the twist angle and $J_\perp$ allow for the identification of many other topologically distinct phases.
This phase diagram shows many topologically distinct phases as a function of commensurability parameter $qL_m$ and the interlayer coupling $J_\perp$. 
The transition into the spiral phase is found to be approximately linear when $J_\perp$ is normalized by the moir\'e area $\sim L_m^2$, which leads to a choice of vertical axis for the phase diagram of the form $J_\perp L_m^2/J_1 a^2$.
The spin spiral phase occurs when $J_\perp/J_1$ is small or the twist angle is large, making $L_m$ too small to impact the interlayer exchange field significantly (Fig.~\ref{fig:phase_diagram}(b)). However, the direction of $\boldsymbol{q}$ is changed from the third neighbors to the moir\'e superlattice vectors. A series of other phases corresponding to  $k\pi-$skyrmion or so-called target skyrmion lattices emerge at integer values of $qL_m/2\pi=k$ ($k\in\mathbb{N}$), when the moir\'e unit cell accommodates an integer number of spin-spiral wavelengths (Figs.~\ref{fig:phase_diagram}(c,e,g)). 
Another family of phases corresponds to nematic skyrmion lattices (Figs.~\ref{fig:phase_diagram}(d,f)), which occur for twist angles that lead to incommensurate values of $q$ and $L_m$. This produces whirling textures within each unit cell and breaks translational symmetry differently for each moir\'e lattice vector as compared to the commensurate case. Therefore, these nematic phases can host coexisting skyrmion and spiral features.

It is worth noting that the distinct emergent phases can be distinguished based on topological features. In particular, the topology of the configuration depends on the distribution of ferromagnetic chains and how they accommodate the local $q$-vector. These chains can form concentric loops corresponding to skyrmionic textures, or cross the borders of the moir\'e unit cell, leading to helical and nematic textures. For non-integer $qL_m$ values, for instance, chains are deformed in such a way as to form horseshoe patterns, or other distinct whirls, locally breaking the described $q-$patches into smaller regions, but preserving some of the global topological features of the spin configuration, such as the total winding $k\pi$. In general, the textures found in between $k\pi$ and $(k+1)\pi-$Skyrmion lattices can have any winding $p\in\mathbb{N}$ with $0\leq p< k+1$. The presence of these topological features points to potential stability under external perturbations.

\textit{Electric field control-} We now address the effect of an external out-of-plane electric field on the ground-state spin configuration in these $k\pi$-skyrmion lattice phases. The strong magnetoelectric coupling in multiferroics is particularly attractive to control skyrmion lattices. An external uniform electric field $\boldsymbol{E}_\text{ext}$ couples to the associated electric polarization $\boldsymbol{P}_{ij}$ generated via the inverse DM. This introduces a magnetoelectric coupling term in the Hamiltonian of the form $H_\text{ME}=-\sum_{\left<ij\right>} \boldsymbol{E}_\text{ext}\cdot\boldsymbol{P}_{ij}.$
Microscopically, this term stems from an inverse DM interaction, leading a term of the form

\begin{equation}
H_\text{ME}=-\lambda E_\perp\sum_{\left<ij\right>}\left(\hat{z}\times\boldsymbol{r}_{ij}\right)\cdot\left(\boldsymbol{S}_i\times\boldsymbol{S}_j\right).
\end{equation}

\begin{figure}
    \centering
    \includegraphics[width=\columnwidth]{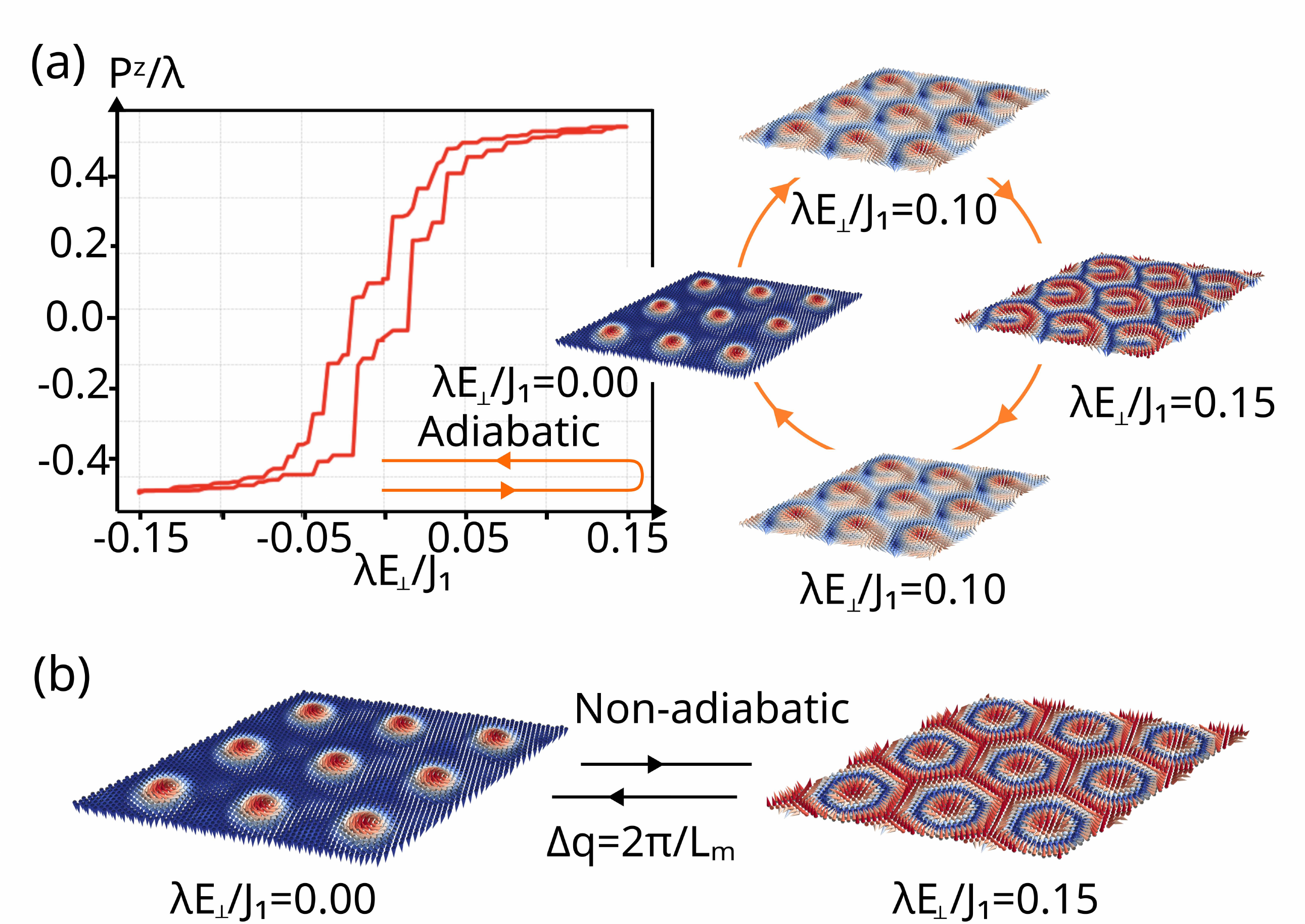}
    \caption{(a) Adiabatic evolution of the skyrmion lattice ground state of twisted NiI$_2$. The polarization along $z$ for the top layer undergoes a hysteresis loop as the field magnitude $\lambda E_\text{ext}/J_1$ is changed along the loop $0\to0.15\to-0.15\to0$ (b) Non-adiabatic evolution of the ground state, resulting in a transition from a skyrmion to a skyrmionium lattice.}
    \label{fig:Electric}
\end{figure}

For $\boldsymbol{E}_\text{ext}\parallel\hat{z}$, this DM term promotes the canting of spins, favoring skyrmion formation, thus enhancing the formation of N\'{e}el-type skyrmion lattices over Bloch-type counterparts. 
Moreover, for a specific moir\'e length scale $L_m$, smaller values of $q$ minimize the contribution from the polarization term, suggesting that the applied electric field can potentially alter the $k\pi$ value of the skyrmion lattice ground state.

For the sake of concreteness, we focus on the skyrmion lattice ground state, although all conclusions drawn here apply to higher order $k\pi$-skyrmion lattices. We focus on two distinct scenarios: subjecting the bilayer system to an external electric field in an adiabatic manner and a non-adiabatic manner. Adiabatically changing the external field strength involves incrementally adjusting the field in small steps while allowing for the spin configuration to relax adiabatically. In contrast, a non-adiabatic change involves applying a
sudden change to the electric field, followed by a full relaxation
of the spin configuration. The results of these calculations are shown in Fig.~\ref{fig:Electric}. 
An adiabatic increase of the electric field preserves the system's topological features, such as the nature of the skyrmion lattice, while merely deforming and shifting the location of the skyrmion center, originating a horseshoe skyrmion texture.
On the other hand, non-adiabatic changes lead to more dramatic effects, altering the $k$ value in a $k\pi$-skyrmion. 
The skyrmion lattice can be transformed into a skyrmionium lattice through a non-adiabatic increase in the electric field.
In this scenario, the external electric field effectively modifies the $q$ of the underlying spin-spiral by $\Delta q =2\pi/L_m$.
Therefore, these results highlight the robustness and magnetoelectric tunability of the topological phases in twisted spin-spiral multiferroics.

\paragraph*{Conclusion-}
We have demonstrated that twisted bilayer NiI$_2$ provides a highly tunable platform for engineering a whole family of topological magnetic phases. Our results establish that twist engineering allows stabilizing a variety of magnetic orders, including skyrmion lattices and nematic phases, thanks to the interplay between interlayer coupling and the moir\'e pattern. The moir\'e-induced modulations stabilize non-uniform $\boldsymbol{q}$-vector configurations, enabling complex magnetic textures and an out-of-plane ferroelectric polarization. The strong magnetoelectric coupling displayed by this twisted multiferroic allows for the manipulation of skyrmion lattice phases via an external electric field. These findings highlight the potential of twisted bilayer NiI$_2$ for exploring unconventional magnetism
and ultimately enabling spintronic devices that exploit electric-field control of topological spin textures.

\paragraph*{Acknowledgements-}
We acknowledge the computational resources provided by the Aalto Science-IT project
and the financial support from the Academy of Finland Projects Nos. 331342, 358088, and 349696,
the Jane and Aatos Erkko Foundation,
and the Finnish Quantum Flagship. 
We thank M. Amini and P. Liljeroth for useful discussions.

\bibliography{biblio}

\end{document}